\documentclass[
reprint,
 amsmath,amssymb,
 aps,
]{revtex4-2}

\usepackage{graphicx}
\usepackage{dcolumn}
\usepackage{bm}

\begin{document}

\preprint{APS/123-QED}

\title{Non-Hermitian Topoelectrical Circuits: Expedient Tools for Topological State Engineering with Gain-Loss Modulation}

\author{Nitish Kumar Gupta}
\email{nitishkg@iitk.ac.in}
\author{Arun M. Jayannavar}

\affiliation{Centre for Lasers \& Photonics, Indian Institute of Technology Kanpur, 208016, India}
\affiliation{Institute of Physics, Bhubaneswar, Odisha 751005, India}

\date{\today}

\begin{abstract}
The congregation of topological quantum and classical systems with the ideas of non-Hermitian physics has generated enormous research interest in the last few years. While the concepts associated to non-trivial topological aspects have provided us an access to the disorder immune states, non-Hermitian physics, which was initially developed within the framework of quantum field theories, has contributed significantly to the study of open quantum systems. Particularly in optics and photonics, the study of non-Hermitian Hamiltonians with balanced loss and gain has resulted in many counter-intuitive phenomena. However, the experimental realization of such systems is challenging, and the need for alternative platforms for testing theoretical propositions and proof of concept demonstrations is widely felt. In this context, active electrical and electronic circuitry has proved to be a prolific alternative and has been receiving increasing attention; mainly, the topoelectric circuits, in many instances, have facilitated the investigation of topological conceptions in conjunction with non-Hermitian physics, beyond the limitations of the condensed matter systems. This article provides a succinct introduction to these non-Hermitian topoelectrical circuits and will also discuss some of the novel physics of topological insulators and semimetals that can be conveniently realized and explored in such configurations.

\end{abstract}
\maketitle

\section{Introduction}

Topology is a branch of mathematics associated with studying the properties of the system that remain invariant under smooth deformations; these properties are generally referred to as the topological invariants, which are integer quantities, potentially useful for system classification. The discovery of the celebrated quantum Hall effect in two-dimensional (2D) electron gas systems brought the topological concepts to the domain of condensed matter physics. The quantum phase transitions in such systems and the appearance of gapless boundary modes were explained with the help of non-trivial topology of the bulk bands, and the associated topological order parameter called the TKNN invariant (which explained the quantization of Hall conductivity). Since then, the domain of topological materials, topological insulators, and semimetals, in particular, has seen many breakthroughs and greatly enriched our understanding of wave transport. The ideas have also been verified in other platforms such as ultracold quantum gases, trapped ions, and superconductors and have opened new frontiers of research~\cite{liu2013detecting,sato2017topological}. Apart from the studies of quantum matter, the numerous application prospects have been guiding the investigations of topological phenomena in classical systems as well, particularly many realizations in photonic and phononic systems~\cite{kim2020recent,gupta2021topological,fleury2016floquet,liu2017model} have been observed in the last decade, resulting in a new paradigm of robust device designs. 

On the other hand, the ideas of non-Hermitian physics have allowed us to look beyond the usual framework of the unitary system evolution of quantum mechanics. In these non-conservative settings, it has been found that counter-intuitive effects such as unidirectional reflectionless phenomena, can be accessed by coupling manipulation between two interacting subsystems~\cite{gu2017unidirectional}. A radical change of perspective about non-Hermitian systems, however, was brought about by Bender and Boettcher~\cite{bender1998real} when they put forward their findings of obtaining real eigenvalue spectra, provided that the system Hamiltonian commutes with the parity-time ($\mathcal{PT}$) operator, where $\mathcal{P}$ is the parity and $\mathcal{T}$ is the time-reversal operator. Realizing these special class of Hamiltonians requires a delicate balance of gain and loss, which gives the photonic realizations an edge over their solid-state electronic counterparts, as optical non-Hermiticities are relatively easier to incorporate. Indeed, the $\mathcal{PT}$-symmetric photonics has experienced rapid developments in just a matter of few years as it facilitates the proliferation of established design principles of photonics to the entire complex permittivity plane~\cite{feng2017non,zhao2019non,yang2021non}. The peculiar characteristics of the associated non-Hermitian singularities and the gain/loss induced phase transition phenomena have been driving the research in this domain~\cite{gupta2020parity,zhou2018dynamical,zhu2021single}. 

Of late, it has been seen that these two emerging arenas are coming together under the aegis of non-Hermitian Topological systems, exhibiting uncanny boundary sensitive phenomena, termed as the breakdown of the conventional bulk-boundary correspondence~\cite{lee2016anomalous} and the associated non-Hermitian skin effect~\cite{okuma2020topological}, which has forced us to adopt alternative viewpoints while dealing with topological aspects of non-Hermitian systems. Indeed, the confluence of the ideas and their reconciliation has not proved to be straightforward- there has been a lot of discussion and debate around defining suitable topological invariants in non-Hermitian systems~\cite{gong2018topological,harari2018topological,leykam2017edge}, which at best can be in dynamic equilibrium only. This has inspired a significant theoretical activity, and neoteric perspectives of defining topological invariants in non-equilibrium settings~\cite{yao2018non,yao2018edge,kunst2018biorthogonal,edvardsson2019non,bergholtz2021exceptional,kunst2019non,koch2020bulk,shen2018topological,xue2020non}. Their experimental verification in fermionic and even in photonic realizations have to face many difficulties on account of the limited control that these systems permit over the hopping strengths, engendering the need for analogous systems with flexible and convenient realization. Electrical and electronic circuitry with a plethora of choices for active and passive lumped components can fulfill these requirements; indeed, in the recent past, these configurations have proved to be a fertile platform for exploring non-Hermitian topological phenomena, enabling us to understand the nuances of unconventional signal transport. These circuits are also known as non-Hermitian topoelectrical circuits and have been used recently for demonstration of spin Hall effect, Haldane model topological states, topological corner modes, topological edge modes in Su-Schrieffer-Heeger (SSH) model setup, Chern insulators, Weyl states, and Fermi-arc surface states~\cite{zhu2018simulating,serra2019observation,ezawa2019electric,ezawa2019non,rafi2021non,haenel2019chern,imhof2018topolectrical,lee2018topolectrical,liu2020gain,goren2018topological,rafi2020topoelectrical,luo2018nodal}. In comparison to any other realization, the inclusion of loss/gain here is just a matter of introducing the lumped resistors between the voltage nodes and ground, which break the Hermiticity of the circuit Laplacian. Furthermore, the circuit realizations offer some coveted functionalities like the incorporation of strong non-linear effects, and nonreciprocity (using varactor diodes)~\cite{hadad2018self}, making them a precious tool for gaining crucial insights and also in deciding the promising research avenues in non-Hermitian topological physics. 
In this article, we will briefly introduce periodic non-Hermitian electrical circuits and will show that they can well emulate the topological properties of native systems. Specifically, we will ponder upon 1D circuits and ascertain their topological characteristics. We will also show that by breaking the Hermiticity condition, the non-Hermitian spectral degeneracies, called the exceptional points (where the system Hamiltonian becomes defective), can be realized in the electrical circuits. Finally, we will discuss the behavior of some specific non-Hermitian topoelectrical circuits in the gain/loss parametric space to fathom their non-trivial topology.

\section{A Paradigmatic Passive Hermitian Topoelectrical Circuit}

At the very outset, we want to briefly discuss the means of experimental characterization of the topoelectrical circuits. The topological edge states in electrical circuits can be identified either by node voltage/branch current measurements or by measuring admittance between neighboring nodes. In our opinion, the approach of measurement of admittance spectra between all the nodes of the circuit is a more reliable approach, which also provides information on the local density of states. This can be accomplished by measuring the S-parameters using a vector network analyzer. With this overview, we begin our sojourn into electrical realizations of topological materials:

\begin{figure}[htbp]
\centering
\fbox{\includegraphics[width=\linewidth]{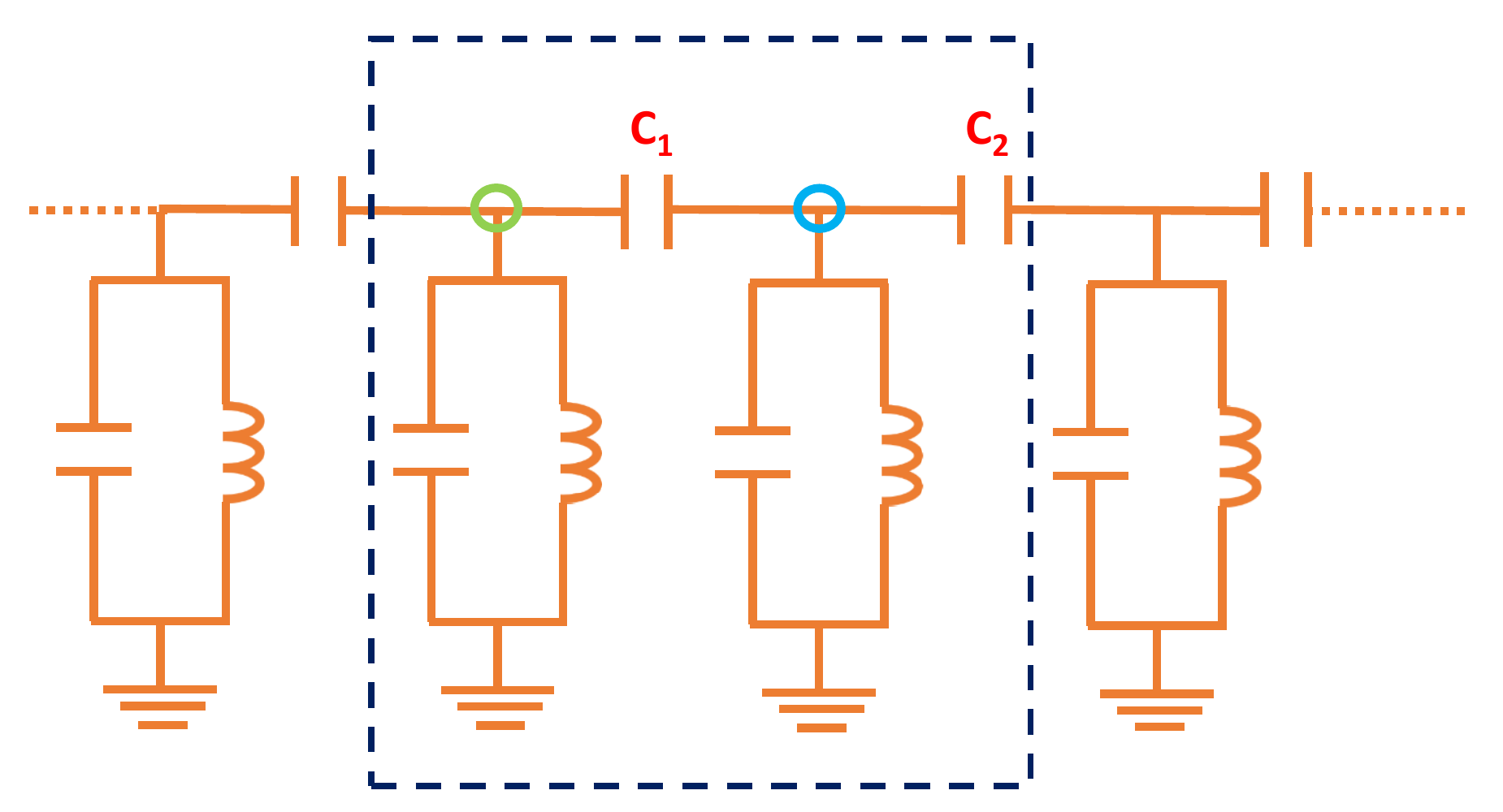}}
\caption{The setup of an archetypal 1D circuit lattice: $C_1$ represents the intra-unitcell capacitive coupling, and $C_2$ represents the inter-unitcell capacitive coupling. The boxed portion of the network depicts a dimerized unit cell.}
\label{fig1}
\end{figure}

The concept was first put forth by Jia Ningyuan et al.~\cite{PhysRevX.5.021031}, where they had presented site- and time-resolved measurements of time-reversal-invariant topological bandstructure in a radio frequency photonic circuit. Since then, such platforms have been utilized to emulate a variety of tight-binding Hamiltonians. Here, we will start our analysis by considering an archetypal periodic electrical circuit- a passive, Hermitian 1D array of LC resonators with capacitive couplings. Such a resonator array can exhibit the topological characteristics of an equivalent SSH model if suitable dimerization is introduced in the network. In this perspective, the unit cell of the resonator array comprises of two sublattice nodes ${N_1}$ and ${N_2}$ with intra-unitcell capacitive coupling ${C_1}$, while the nearest neighbor inter-unitcell capacitive coupling is represented by ${C_2}$. The grounding of the nodes is provided by a parallel $LC$ circuit. The configuration under consideration is represented in Fig. 1. By employing Kirchhoff's laws, we can deduce the circuit Laplacian (or equivalently the admittance matrix) for this periodic network, which can be written in a simplified form as:

\begin{equation}
    \boldsymbol{J}(q)= -[C_1 + C_2 \cos(q)] \begin{pmatrix}
0 & 1\\
1 & 0
\end{pmatrix} - C_2 \sin(q) \begin{pmatrix}
0 & -i\\
i & 0
\end{pmatrix}
\end{equation}

where the matrices in question are the Pauli matrices $\boldsymbol{\sigma_x}$, and  $\boldsymbol{\sigma_y}$, in the sublattice space, and $q$ is the Bloch momentum which links a given unit cell with its neighbor. The periodicity of the circuit permits the use of the Bloch theorem to find wave functions. Evidently, this equation is very similar to the tight-binding Hamiltonians obtained in 1D condensed matter systems; thus, the admittance matrix can be interpreted as the system Hamiltonian. In the absence of any dissipative circuit element, the above circuit Laplacian is Hermitian, and hence, it would possess real admittance eigenvalues ${y(q)}$. The admittance eigenvalue spectra for the circuit Laplacian corresponding to Eq. (1) have been calculated and plotted in Fig. (2) for three representative cases based on the coupling ratio. 
The topological invariant for the admittance bandstructure can be  extracted from the corresponding bulk eigenfunctions ${\psi(q)}$, using the standard definition of winding number:

\begin{equation}
    \boldsymbol{\mathcal{W}}= -\frac{1}{\pi}  \oint i\left\langle \psi(q)  \middle| \nabla_{q} \middle| \psi(q) \right\rangle dq 
\end{equation}

These winding numbers are mentioned alongside the bandstructures of Fig. 2, which demonstrate the existence of non-trivial topological characteristics in the admittance bandstructure for the case of $C_1<C_2$. The nonzero value of the winding number, in this case, suggests that a transition to a trivial case would require a closure of the spectral gap of the admittance bandstructure.

\begin{figure}[htbp]
\centering
\fbox{\includegraphics[width=\linewidth]{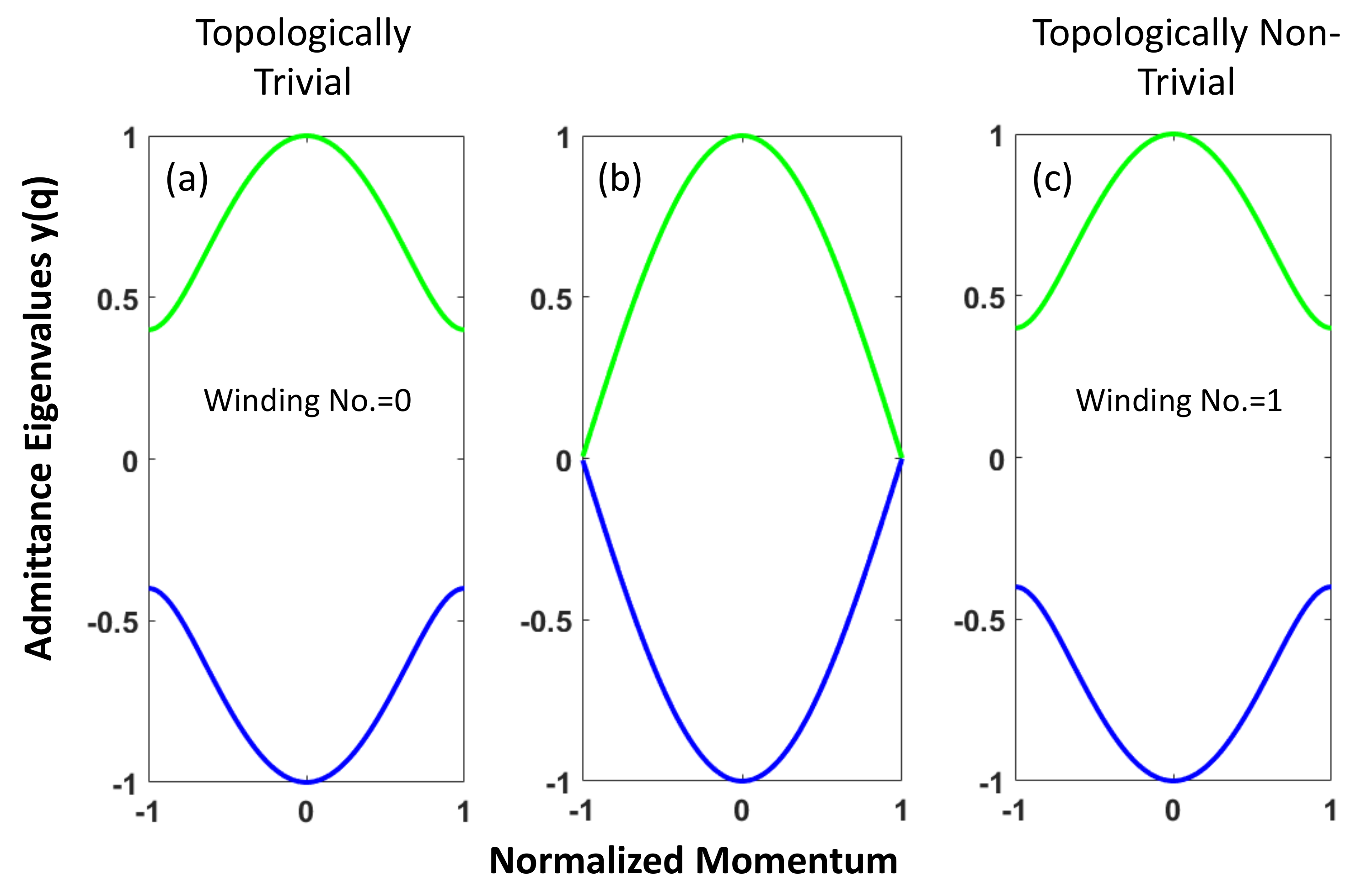}}
\caption{Calculated admittance eigenvalue spectra for the circuit SSH realization of Fig. 1 : (a) $C_1>C_2$, (b) $C_1=C_2$, (c) $C_1<C_2$.}
\label{fig2}
\end{figure}

\section{Breaking the Hermiticity of Circuit Laplacian}

As mentioned before, significant research effort has been devoted in the last few years to understand the topological phases and the associated aspects in the non-Hermitian systems. In this context, the relative ease of inclusion of non-conservative system parameters in electrical circuits has proved to be one of the major impetus behind the rapid rise in research works discussing topological characteristics in electrical circuits. This section will begin by discussing the means to incorporate non-Hermiticity in the circuit Laplacian and some of the peculiarities it brings to the bandstructure. Then we will move on to provide a brief account of some recently published research articles on topoelectrical circuits, which have studied the effects of non-Hermiticity on popular topological platforms.

The practical means of incorporating loss in the topoelectrical circuit is by the introduction of a resistive element. Incorporation of gain, however, necessitates the use of active circuit elements. One very popular circuit which can realize an arbitrary value of gain (as an effective negative resistance) is the op-amp-based Negative Impedance Converter with current inversion (NIC) circuit. This circuit has been recently employed by Shou Liu et al. to study gain (loss) induced topological states in a finite unit cell network~\cite{liu2020gain}. 

When we introduce gain-loss in the system in a balanced manner, the system respects $\mathcal{PT}$ symmetry, and the possibility of obtaining exceptional points arises. To highlight these prospects, we have modified the topoelectrical lattice of Fig. 1 and introduced identical positive and negative resistances (or conductances) at alternate sites in parallel to the LC circuit.  Such a construct is equivalent to introducing onsite imaginary potentials and offers a direct correspondence with already discussed theoretical aspects. The value of the resistance now forms a parameter of the $\mathcal{PT}$ symmetric system, which can be suitably changed to derive different regimes of operation. We have observed that for small values of the non-Hermiticity parameter, the system remains in the unbroken $\mathcal{PT}$ phase while beyond a certain value of the parameter, the system makes a spontaneous transition to the broken $\mathcal{PT}$ phase and the exceptional points show up in the admittance eigenvalue spectra. These aspects have been demonstrated in Fig. 3, where we can observe the existence of exceptional points in the real and imaginary eigenvalue spectra in all the three configurations corresponding to Fig. 2.

For a system-specific discussion on the topological aspects of non-Hermitian electrical realizations, we focus on two of the recent reports and the methodologies adopted in them:

The first work that we will be discussing is the reference~\cite{liu2020gain} where the authors have analyzed and demonstrated how gain and loss could be utilized to induce topological insulating phase in a circuit realization of the SSH model. The report works with a 1D non-Hermitian circuit array of parallel LC circuits, where positive and negative resistors are arranged in a pair of resonator dimers. The network arrangement is analyzed by writing the frequency-dependent circuit Laplacian matrix. The Laplacian satisfies the pseudo-Hermiticity and pseudo-anti-Hermiticity, giving a chiral nature to the admittance spectra. For different combinations of gain and loss values, the circuit exhibits four different phases,  which were identified by the crossings between the bulk bands. The topological characteristics of these phases have been studied by defining a normalized global Berry phase which serves as the topological invariant
\begin{equation}
    W=  \sum_{s} \frac{i}{4\pi} \oint\langle\langle \psi_{B,s}|\nabla_k|\psi_{B,s} \rangle~dq
\end{equation}
which reveals that topological characteristics persist in only three of the configurations. Experimental demonstrations were also conducted in a ten-unit cell network where edge states have been observed by noticing the pronounced impedance peaks. 

\begin{figure}[htbp]
\centering
\fbox{\includegraphics[width=\linewidth]{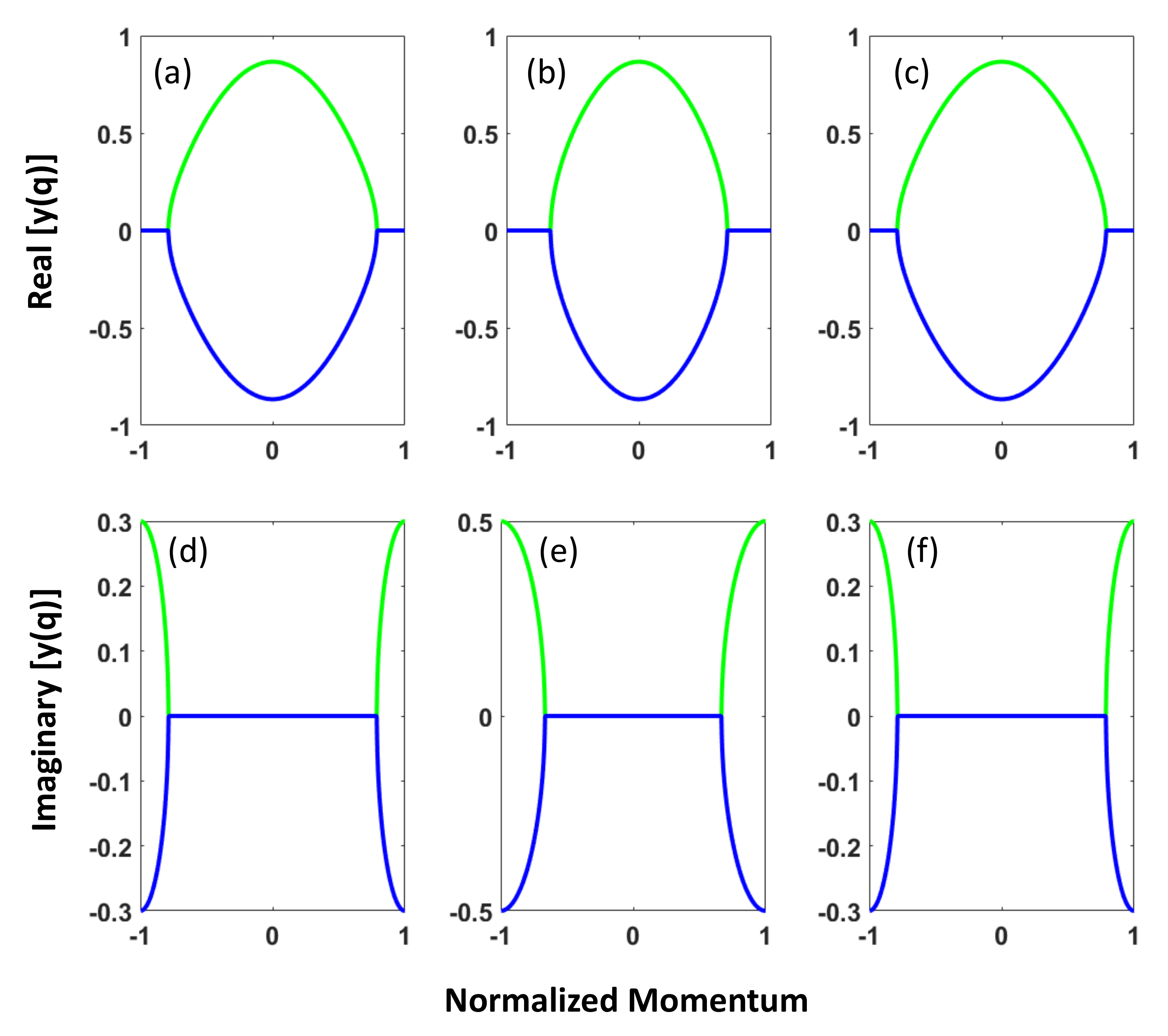}}
\caption{Calculated real and imaginary admittance bandstructure for the non-Hermitian circuit SSH realization containing balanced loss-gain elements: (a),(d) $C_1>C_2$; (b),(e) $C_1=C_2$; (c),(f) $C_1<C_2$. The spectra are plotted for above the critical value of the non-Hermiticity parameter.}
\label{fig3}
\end{figure}

Another work that we will ponder upon is the reference~\cite{ezawa2019electric}, where the author has analyzed in detail the circuit realization of the non-Hermitian  Haldane model (by making the spin-orbit interaction itself to be non-Hermitian). Each node of the 2D lattice consists of a parallel RLC circuit, while the non-Hermitian Haldane interaction has been realized using capacitors and op-amps (in the negative impedance configuration). Such a spin-orbit term yields a complex Dirac mass at the high symmetry points. The topological phase transition in the configuration can be ascertained by defining a topological invariant called the non-Hermitian Chern number, defined as

\begin{equation}
    W=   \frac{1}{2\pi} \oint F(\boldsymbol{k}) d^2\boldsymbol{k}
\end{equation}

where $F(\boldsymbol{k})$ is the non-Hermitian Berry curvature given by $F(\boldsymbol{k})=\nabla\times A(\boldsymbol{k})$. Experimentally, the phase transition can be detected by measuring the change of impedance. The author has noted that in this circuit, the left-going and right-going chiral edge modes
can be distinguished by the phase of the two-point impedance. In other words, the phase of the impedance acquires a dynamical degree of freedom in the non-Hermitian realizations.

We chose to discuss these two realizations as they pertain to the non-Hermitian electric circuit analog of the two of the most celebrated condensed matter topological systems. As stated earlier, the circuit realizations have been employed in many other instances and have been a front-runner in pushing the boundaries of our understanding of topological physics at large~\cite{PhysRevLett.124.046401,PhysRevB.100.165419}.

\section{Conclusion}

Beginning with a succinct introduction of some of the peculiar topological aspects of non-Hermitian systems, we went on to dwell upon one particular classical realization of topological quantum matter, namely the topoelectrical circuits. These realizations have gained prominence in the past few years as relatively less demanding platforms to study the topological characteristics in non-Hermitian settings. In this context, a systematic account of topoelectrical circuits has been provided by first defining an archetypal Hermitian LC ladder network and understanding the origins of topological characteristics in it. Then we moved on to break the Hermiticity of the associated circuit Laplacian by adding balanced gain and loss, which engenders the possibility of observing exceptional points in the bulk bandstructure. These exceptional points serve as the boundary between different regimes of operations. Finally, the topological aspects of the non-Hermitian topoelectrical circuits have been discussed in consonance with the perspective provided by two of the recent realizations of the SSH and Haldane model. We hope to have convinced the reader about a justified correspondence between the topoelectrical circuits and condensed matter systems, which permits a direct mapping of physical concepts. Indeed, most of the physics of tight-binding Hamiltonians can be mapped to the corresponding electrical realizations, offering us a prolific means for proof of concept demonstrations and further explorations of the domain.

\section{Author Information}
Nitish Kumar Gupta (email-nitishkg@iitk.ac.in) is with Centre for Lasers \& Photonics, Indian Institute of Technology Kanpur, 208016, India.  A. M. Jayannavar (email-jayan@iopb.res.in) is a J C Bose national fellow and a senior scientist currently associated with Bhaurao Kakatkar College, Belgaum, India.

\section{Acknowledgement}

A. M. Jayannavar acknowledges DST, India for J C Bose fellowship.

\bibliography{references_aps}

\providecommand{\noopsort}[1]{}\providecommand{\singleletter}[1]{#1}%
\begin{thebibliography}{44}%
\makeatletter
\providecommand \@ifxundefined [1]{%
 \@ifx{#1\undefined}
}%
\providecommand \@ifnum [1]{%
 \ifnum #1\expandafter \@firstoftwo
 \else \expandafter \@secondoftwo
 \fi
}%
\providecommand \@ifx [1]{%
 \ifx #1\expandafter \@firstoftwo
 \else \expandafter \@secondoftwo
 \fi
}%
\providecommand \natexlab [1]{#1}%
\providecommand \enquote  [1]{``#1''}%
\providecommand \bibnamefont  [1]{#1}%
\providecommand \bibfnamefont [1]{#1}%
\providecommand \citenamefont [1]{#1}%
\providecommand \href@noop [0]{\@secondoftwo}%
\providecommand \href [0]{\begingroup \@sanitize@url \@href}%
\providecommand \@href[1]{\@@startlink{#1}\@@href}%
\providecommand \@@href[1]{\endgroup#1\@@endlink}%
\providecommand \@sanitize@url [0]{\catcode `\\12\catcode `\$12\catcode
  `\&12\catcode `\#12\catcode `\^12\catcode `\_12\catcode `\%12\relax}%
\providecommand \@@startlink[1]{}%
\providecommand \@@endlink[0]{}%
\providecommand \url  [0]{\begingroup\@sanitize@url \@url }%
\providecommand \@url [1]{\endgroup\@href {#1}{\urlprefix }}%
\providecommand \urlprefix  [0]{URL }%
\providecommand \Eprint [0]{\href }%
\providecommand \doibase [0]{https://doi.org/}%
\providecommand \selectlanguage [0]{\@gobble}%
\providecommand \bibinfo  [0]{\@secondoftwo}%
\providecommand \bibfield  [0]{\@secondoftwo}%
\providecommand \translation [1]{[#1]}%
\providecommand \BibitemOpen [0]{}%
\providecommand \bibitemStop [0]{}%
\providecommand \bibitemNoStop [0]{.\EOS\space}%
\providecommand \EOS [0]{\spacefactor3000\relax}%
\providecommand \BibitemShut  [1]{\csname bibitem#1\endcsname}%
\let\auto@bib@innerbib\@empty
\bibitem [{\citenamefont {Liu}\ \emph {et~al.}(2013)\citenamefont {Liu},
  \citenamefont {Law}, \citenamefont {Ng},\ and\ \citenamefont
  {Lee}}]{liu2013detecting}%
  \BibitemOpen
  \bibfield  {author} {\bibinfo {author} {\bibfnamefont {X.-J.}\ \bibnamefont
  {Liu}}, \bibinfo {author} {\bibfnamefont {K.-T.}\ \bibnamefont {Law}},
  \bibinfo {author} {\bibfnamefont {T.-K.}\ \bibnamefont {Ng}},\ and\ \bibinfo
  {author} {\bibfnamefont {P.~A.}\ \bibnamefont {Lee}},\ }\bibfield  {title}
  {\bibinfo {title} {Detecting topological phases in cold atoms},\ }\href@noop
  {} {\bibfield  {journal} {\bibinfo  {journal} {Physical review letters}\
  }\textbf {\bibinfo {volume} {111}},\ \bibinfo {pages} {120402} (\bibinfo
  {year} {2013})}\BibitemShut {NoStop}%
\bibitem [{\citenamefont {Sato}\ and\ \citenamefont
  {Ando}(2017)}]{sato2017topological}%
  \BibitemOpen
  \bibfield  {author} {\bibinfo {author} {\bibfnamefont {M.}~\bibnamefont
  {Sato}}\ and\ \bibinfo {author} {\bibfnamefont {Y.}~\bibnamefont {Ando}},\
  }\bibfield  {title} {\bibinfo {title} {Topological superconductors: a
  review},\ }\href@noop {} {\bibfield  {journal} {\bibinfo  {journal} {Reports
  on Progress in Physics}\ }\textbf {\bibinfo {volume} {80}},\ \bibinfo {pages}
  {076501} (\bibinfo {year} {2017})}\BibitemShut {NoStop}%
\bibitem [{\citenamefont {Kim}\ \emph {et~al.}(2020)\citenamefont {Kim},
  \citenamefont {Jacob},\ and\ \citenamefont {Rho}}]{kim2020recent}%
  \BibitemOpen
  \bibfield  {author} {\bibinfo {author} {\bibfnamefont {M.}~\bibnamefont
  {Kim}}, \bibinfo {author} {\bibfnamefont {Z.}~\bibnamefont {Jacob}},\ and\
  \bibinfo {author} {\bibfnamefont {J.}~\bibnamefont {Rho}},\ }\bibfield
  {title} {\bibinfo {title} {Recent advances in 2d, 3d and higher-order
  topological photonics},\ }\href@noop {} {\bibfield  {journal} {\bibinfo
  {journal} {Light: Science \& Applications}\ }\textbf {\bibinfo {volume}
  {9}},\ \bibinfo {pages} {1} (\bibinfo {year} {2020})}\BibitemShut {NoStop}%
\bibitem [{\citenamefont {Gupta}\ and\ \citenamefont
  {Jayannavar}(2021)}]{gupta2021topological}%
  \BibitemOpen
  \bibfield  {author} {\bibinfo {author} {\bibfnamefont {N.~K.}\ \bibnamefont
  {Gupta}}\ and\ \bibinfo {author} {\bibfnamefont {A.~M.}\ \bibnamefont
  {Jayannavar}},\ }\bibfield  {title} {\bibinfo {title} {Topological photonic
  systems: Virtuous platforms to study topological quantum matter},\
  }\href@noop {} {\bibfield  {journal} {\bibinfo  {journal} {arXiv preprint
  arXiv:2108.05845}\ } (\bibinfo {year} {2021})}\BibitemShut {NoStop}%
\bibitem [{\citenamefont {Fleury}\ \emph {et~al.}(2016)\citenamefont {Fleury},
  \citenamefont {Khanikaev},\ and\ \citenamefont {Alu}}]{fleury2016floquet}%
  \BibitemOpen
  \bibfield  {author} {\bibinfo {author} {\bibfnamefont {R.}~\bibnamefont
  {Fleury}}, \bibinfo {author} {\bibfnamefont {A.~B.}\ \bibnamefont
  {Khanikaev}},\ and\ \bibinfo {author} {\bibfnamefont {A.}~\bibnamefont
  {Alu}},\ }\bibfield  {title} {\bibinfo {title} {Floquet topological
  insulators for sound},\ }\href@noop {} {\bibfield  {journal} {\bibinfo
  {journal} {Nature communications}\ }\textbf {\bibinfo {volume} {7}},\
  \bibinfo {pages} {1} (\bibinfo {year} {2016})}\BibitemShut {NoStop}%
\bibitem [{\citenamefont {Liu}\ \emph {et~al.}(2017)\citenamefont {Liu},
  \citenamefont {Xu}, \citenamefont {Zhang},\ and\ \citenamefont
  {Duan}}]{liu2017model}%
  \BibitemOpen
  \bibfield  {author} {\bibinfo {author} {\bibfnamefont {Y.}~\bibnamefont
  {Liu}}, \bibinfo {author} {\bibfnamefont {Y.}~\bibnamefont {Xu}}, \bibinfo
  {author} {\bibfnamefont {S.-C.}\ \bibnamefont {Zhang}},\ and\ \bibinfo
  {author} {\bibfnamefont {W.}~\bibnamefont {Duan}},\ }\bibfield  {title}
  {\bibinfo {title} {Model for topological phononics and phonon diode},\
  }\href@noop {} {\bibfield  {journal} {\bibinfo  {journal} {Physical Review
  B}\ }\textbf {\bibinfo {volume} {96}},\ \bibinfo {pages} {064106} (\bibinfo
  {year} {2017})}\BibitemShut {NoStop}%
\bibitem [{\citenamefont {Gu}\ \emph {et~al.}(2017)\citenamefont {Gu},
  \citenamefont {Bai}, \citenamefont {Zhang}, \citenamefont {Jin},
  \citenamefont {Zhang}, \citenamefont {Zhang},\ and\ \citenamefont
  {Lee}}]{gu2017unidirectional}%
  \BibitemOpen
  \bibfield  {author} {\bibinfo {author} {\bibfnamefont {X.}~\bibnamefont
  {Gu}}, \bibinfo {author} {\bibfnamefont {R.}~\bibnamefont {Bai}}, \bibinfo
  {author} {\bibfnamefont {C.}~\bibnamefont {Zhang}}, \bibinfo {author}
  {\bibfnamefont {X.~R.}\ \bibnamefont {Jin}}, \bibinfo {author} {\bibfnamefont
  {Y.~Q.}\ \bibnamefont {Zhang}}, \bibinfo {author} {\bibfnamefont
  {S.}~\bibnamefont {Zhang}},\ and\ \bibinfo {author} {\bibfnamefont {Y.~P.}\
  \bibnamefont {Lee}},\ }\bibfield  {title} {\bibinfo {title} {Unidirectional
  reflectionless propagation in a non-ideal parity-time metasurface based on
  far field coupling},\ }\href@noop {} {\bibfield  {journal} {\bibinfo
  {journal} {Optics express}\ }\textbf {\bibinfo {volume} {25}},\ \bibinfo
  {pages} {11778} (\bibinfo {year} {2017})}\BibitemShut {NoStop}%
\bibitem [{\citenamefont {Bender}\ and\ \citenamefont
  {Boettcher}(1998)}]{bender1998real}%
  \BibitemOpen
  \bibfield  {author} {\bibinfo {author} {\bibfnamefont {C.~M.}\ \bibnamefont
  {Bender}}\ and\ \bibinfo {author} {\bibfnamefont {S.}~\bibnamefont
  {Boettcher}},\ }\bibfield  {title} {\bibinfo {title} {Real spectra in
  non-hermitian hamiltonians having p t symmetry},\ }\href@noop {} {\bibfield
  {journal} {\bibinfo  {journal} {Physical Review Letters}\ }\textbf {\bibinfo
  {volume} {80}},\ \bibinfo {pages} {5243} (\bibinfo {year}
  {1998})}\BibitemShut {NoStop}%
\bibitem [{\citenamefont {Feng}\ \emph {et~al.}(2017)\citenamefont {Feng},
  \citenamefont {El-Ganainy},\ and\ \citenamefont {Ge}}]{feng2017non}%
  \BibitemOpen
  \bibfield  {author} {\bibinfo {author} {\bibfnamefont {L.}~\bibnamefont
  {Feng}}, \bibinfo {author} {\bibfnamefont {R.}~\bibnamefont {El-Ganainy}},\
  and\ \bibinfo {author} {\bibfnamefont {L.}~\bibnamefont {Ge}},\ }\bibfield
  {title} {\bibinfo {title} {Non-hermitian photonics based on parity--time
  symmetry},\ }\href@noop {} {\bibfield  {journal} {\bibinfo  {journal} {Nature
  Photonics}\ }\textbf {\bibinfo {volume} {11}},\ \bibinfo {pages} {752}
  (\bibinfo {year} {2017})}\BibitemShut {NoStop}%
\bibitem [{\citenamefont {Zhao}\ \emph {et~al.}(2019)\citenamefont {Zhao},
  \citenamefont {Qiao}, \citenamefont {Wu}, \citenamefont {Midya},
  \citenamefont {Longhi},\ and\ \citenamefont {Feng}}]{zhao2019non}%
  \BibitemOpen
  \bibfield  {author} {\bibinfo {author} {\bibfnamefont {H.}~\bibnamefont
  {Zhao}}, \bibinfo {author} {\bibfnamefont {X.}~\bibnamefont {Qiao}}, \bibinfo
  {author} {\bibfnamefont {T.}~\bibnamefont {Wu}}, \bibinfo {author}
  {\bibfnamefont {B.}~\bibnamefont {Midya}}, \bibinfo {author} {\bibfnamefont
  {S.}~\bibnamefont {Longhi}},\ and\ \bibinfo {author} {\bibfnamefont
  {L.}~\bibnamefont {Feng}},\ }\bibfield  {title} {\bibinfo {title}
  {Non-hermitian topological light steering},\ }\href@noop {} {\bibfield
  {journal} {\bibinfo  {journal} {Science}\ }\textbf {\bibinfo {volume}
  {365}},\ \bibinfo {pages} {1163} (\bibinfo {year} {2019})}\BibitemShut
  {NoStop}%
\bibitem [{\citenamefont {Yang}\ \emph {et~al.}(2021)\citenamefont {Yang},
  \citenamefont {Hwang}, \citenamefont {Doiron},\ and\ \citenamefont
  {Naik}}]{yang2021non}%
  \BibitemOpen
  \bibfield  {author} {\bibinfo {author} {\bibfnamefont {F.}~\bibnamefont
  {Yang}}, \bibinfo {author} {\bibfnamefont {A.}~\bibnamefont {Hwang}},
  \bibinfo {author} {\bibfnamefont {C.}~\bibnamefont {Doiron}},\ and\ \bibinfo
  {author} {\bibfnamefont {G.~V.}\ \bibnamefont {Naik}},\ }\bibfield  {title}
  {\bibinfo {title} {Non-hermitian metasurfaces for the best of plasmonics and
  dielectrics},\ }\href@noop {} {\bibfield  {journal} {\bibinfo  {journal}
  {Optical Materials Express}\ }\textbf {\bibinfo {volume} {11}},\ \bibinfo
  {pages} {2326} (\bibinfo {year} {2021})}\BibitemShut {NoStop}%
\bibitem [{\citenamefont {Gupta}\ \emph {et~al.}(2020)\citenamefont {Gupta},
  \citenamefont {Zou}, \citenamefont {Zhu}, \citenamefont {Lu}, \citenamefont
  {Zhang}, \citenamefont {Liu},\ and\ \citenamefont {Chen}}]{gupta2020parity}%
  \BibitemOpen
  \bibfield  {author} {\bibinfo {author} {\bibfnamefont {S.~K.}\ \bibnamefont
  {Gupta}}, \bibinfo {author} {\bibfnamefont {Y.}~\bibnamefont {Zou}}, \bibinfo
  {author} {\bibfnamefont {X.-Y.}\ \bibnamefont {Zhu}}, \bibinfo {author}
  {\bibfnamefont {M.-H.}\ \bibnamefont {Lu}}, \bibinfo {author} {\bibfnamefont
  {L.-J.}\ \bibnamefont {Zhang}}, \bibinfo {author} {\bibfnamefont {X.-P.}\
  \bibnamefont {Liu}},\ and\ \bibinfo {author} {\bibfnamefont {Y.-F.}\
  \bibnamefont {Chen}},\ }\bibfield  {title} {\bibinfo {title} {Parity-time
  symmetry in non-hermitian complex optical media},\ }\href@noop {} {\bibfield
  {journal} {\bibinfo  {journal} {Advanced Materials}\ }\textbf {\bibinfo
  {volume} {32}},\ \bibinfo {pages} {1903639} (\bibinfo {year}
  {2020})}\BibitemShut {NoStop}%
\bibitem [{\citenamefont {Zhou}\ \emph {et~al.}(2018)\citenamefont {Zhou},
  \citenamefont {Wang}, \citenamefont {Wang},\ and\ \citenamefont
  {Gong}}]{zhou2018dynamical}%
  \BibitemOpen
  \bibfield  {author} {\bibinfo {author} {\bibfnamefont {L.}~\bibnamefont
  {Zhou}}, \bibinfo {author} {\bibfnamefont {Q.-h.}\ \bibnamefont {Wang}},
  \bibinfo {author} {\bibfnamefont {H.}~\bibnamefont {Wang}},\ and\ \bibinfo
  {author} {\bibfnamefont {J.}~\bibnamefont {Gong}},\ }\bibfield  {title}
  {\bibinfo {title} {Dynamical quantum phase transitions in non-hermitian
  lattices},\ }\href@noop {} {\bibfield  {journal} {\bibinfo  {journal}
  {Physical Review A}\ }\textbf {\bibinfo {volume} {98}},\ \bibinfo {pages}
  {022129} (\bibinfo {year} {2018})}\BibitemShut {NoStop}%
\bibitem [{\citenamefont {Zhu}\ \emph {et~al.}(2021)\citenamefont {Zhu},
  \citenamefont {Wang}, \citenamefont {Zeng}, \citenamefont {Wang},\ and\
  \citenamefont {Chong}}]{zhu2021single}%
  \BibitemOpen
  \bibfield  {author} {\bibinfo {author} {\bibfnamefont {B.}~\bibnamefont
  {Zhu}}, \bibinfo {author} {\bibfnamefont {Q.}~\bibnamefont {Wang}}, \bibinfo
  {author} {\bibfnamefont {Y.}~\bibnamefont {Zeng}}, \bibinfo {author}
  {\bibfnamefont {Q.~J.}\ \bibnamefont {Wang}},\ and\ \bibinfo {author}
  {\bibfnamefont {Y.}~\bibnamefont {Chong}},\ }\bibfield  {title} {\bibinfo
  {title} {Single-mode lasing based on pt-breaking of two-dimensional photonic
  higher-order topological insulator},\ }\href@noop {} {\bibfield  {journal}
  {\bibinfo  {journal} {arXiv preprint arXiv:2107.02989}\ } (\bibinfo {year}
  {2021})}\BibitemShut {NoStop}%
\bibitem [{\citenamefont {Lee}(2016)}]{lee2016anomalous}%
  \BibitemOpen
  \bibfield  {author} {\bibinfo {author} {\bibfnamefont {T.~E.}\ \bibnamefont
  {Lee}},\ }\bibfield  {title} {\bibinfo {title} {Anomalous edge state in a
  non-hermitian lattice},\ }\href@noop {} {\bibfield  {journal} {\bibinfo
  {journal} {Physical review letters}\ }\textbf {\bibinfo {volume} {116}},\
  \bibinfo {pages} {133903} (\bibinfo {year} {2016})}\BibitemShut {NoStop}%
\bibitem [{\citenamefont {Okuma}\ \emph {et~al.}(2020)\citenamefont {Okuma},
  \citenamefont {Kawabata}, \citenamefont {Shiozaki},\ and\ \citenamefont
  {Sato}}]{okuma2020topological}%
  \BibitemOpen
  \bibfield  {author} {\bibinfo {author} {\bibfnamefont {N.}~\bibnamefont
  {Okuma}}, \bibinfo {author} {\bibfnamefont {K.}~\bibnamefont {Kawabata}},
  \bibinfo {author} {\bibfnamefont {K.}~\bibnamefont {Shiozaki}},\ and\
  \bibinfo {author} {\bibfnamefont {M.}~\bibnamefont {Sato}},\ }\bibfield
  {title} {\bibinfo {title} {Topological origin of non-hermitian skin
  effects},\ }\href@noop {} {\bibfield  {journal} {\bibinfo  {journal}
  {Physical review letters}\ }\textbf {\bibinfo {volume} {124}},\ \bibinfo
  {pages} {086801} (\bibinfo {year} {2020})}\BibitemShut {NoStop}%
\bibitem [{\citenamefont {Gong}\ \emph {et~al.}(2018)\citenamefont {Gong},
  \citenamefont {Ashida}, \citenamefont {Kawabata}, \citenamefont {Takasan},
  \citenamefont {Higashikawa},\ and\ \citenamefont
  {Ueda}}]{gong2018topological}%
  \BibitemOpen
  \bibfield  {author} {\bibinfo {author} {\bibfnamefont {Z.}~\bibnamefont
  {Gong}}, \bibinfo {author} {\bibfnamefont {Y.}~\bibnamefont {Ashida}},
  \bibinfo {author} {\bibfnamefont {K.}~\bibnamefont {Kawabata}}, \bibinfo
  {author} {\bibfnamefont {K.}~\bibnamefont {Takasan}}, \bibinfo {author}
  {\bibfnamefont {S.}~\bibnamefont {Higashikawa}},\ and\ \bibinfo {author}
  {\bibfnamefont {M.}~\bibnamefont {Ueda}},\ }\bibfield  {title} {\bibinfo
  {title} {Topological phases of non-hermitian systems},\ }\href@noop {}
  {\bibfield  {journal} {\bibinfo  {journal} {Physical Review X}\ }\textbf
  {\bibinfo {volume} {8}},\ \bibinfo {pages} {031079} (\bibinfo {year}
  {2018})}\BibitemShut {NoStop}%
\bibitem [{\citenamefont {Harari}\ \emph {et~al.}(2018)\citenamefont {Harari},
  \citenamefont {Bandres}, \citenamefont {Lumer}, \citenamefont {Rechtsman},
  \citenamefont {Chong}, \citenamefont {Khajavikhan}, \citenamefont
  {Christodoulides},\ and\ \citenamefont {Segev}}]{harari2018topological}%
  \BibitemOpen
  \bibfield  {author} {\bibinfo {author} {\bibfnamefont {G.}~\bibnamefont
  {Harari}}, \bibinfo {author} {\bibfnamefont {M.~A.}\ \bibnamefont {Bandres}},
  \bibinfo {author} {\bibfnamefont {Y.}~\bibnamefont {Lumer}}, \bibinfo
  {author} {\bibfnamefont {M.~C.}\ \bibnamefont {Rechtsman}}, \bibinfo {author}
  {\bibfnamefont {Y.~D.}\ \bibnamefont {Chong}}, \bibinfo {author}
  {\bibfnamefont {M.}~\bibnamefont {Khajavikhan}}, \bibinfo {author}
  {\bibfnamefont {D.~N.}\ \bibnamefont {Christodoulides}},\ and\ \bibinfo
  {author} {\bibfnamefont {M.}~\bibnamefont {Segev}},\ }\bibfield  {title}
  {\bibinfo {title} {Topological insulator laser: theory},\ }\href@noop {}
  {\bibfield  {journal} {\bibinfo  {journal} {Science}\ }\textbf {\bibinfo
  {volume} {359}} (\bibinfo {year} {2018})}\BibitemShut {NoStop}%
\bibitem [{\citenamefont {Leykam}\ \emph {et~al.}(2017)\citenamefont {Leykam},
  \citenamefont {Bliokh}, \citenamefont {Huang}, \citenamefont {Chong},\ and\
  \citenamefont {Nori}}]{leykam2017edge}%
  \BibitemOpen
  \bibfield  {author} {\bibinfo {author} {\bibfnamefont {D.}~\bibnamefont
  {Leykam}}, \bibinfo {author} {\bibfnamefont {K.~Y.}\ \bibnamefont {Bliokh}},
  \bibinfo {author} {\bibfnamefont {C.}~\bibnamefont {Huang}}, \bibinfo
  {author} {\bibfnamefont {Y.~D.}\ \bibnamefont {Chong}},\ and\ \bibinfo
  {author} {\bibfnamefont {F.}~\bibnamefont {Nori}},\ }\bibfield  {title}
  {\bibinfo {title} {Edge modes, degeneracies, and topological numbers in
  non-hermitian systems},\ }\href@noop {} {\bibfield  {journal} {\bibinfo
  {journal} {Physical review letters}\ }\textbf {\bibinfo {volume} {118}},\
  \bibinfo {pages} {040401} (\bibinfo {year} {2017})}\BibitemShut {NoStop}%
\bibitem [{\citenamefont {Yao}\ \emph {et~al.}(2018)\citenamefont {Yao},
  \citenamefont {Song},\ and\ \citenamefont {Wang}}]{yao2018non}%
  \BibitemOpen
  \bibfield  {author} {\bibinfo {author} {\bibfnamefont {S.}~\bibnamefont
  {Yao}}, \bibinfo {author} {\bibfnamefont {F.}~\bibnamefont {Song}},\ and\
  \bibinfo {author} {\bibfnamefont {Z.}~\bibnamefont {Wang}},\ }\bibfield
  {title} {\bibinfo {title} {Non-hermitian chern bands},\ }\href@noop {}
  {\bibfield  {journal} {\bibinfo  {journal} {Physical review letters}\
  }\textbf {\bibinfo {volume} {121}},\ \bibinfo {pages} {136802} (\bibinfo
  {year} {2018})}\BibitemShut {NoStop}%
\bibitem [{\citenamefont {Yao}\ and\ \citenamefont {Wang}(2018)}]{yao2018edge}%
  \BibitemOpen
  \bibfield  {author} {\bibinfo {author} {\bibfnamefont {S.}~\bibnamefont
  {Yao}}\ and\ \bibinfo {author} {\bibfnamefont {Z.}~\bibnamefont {Wang}},\
  }\bibfield  {title} {\bibinfo {title} {Edge states and topological invariants
  of non-hermitian systems},\ }\href@noop {} {\bibfield  {journal} {\bibinfo
  {journal} {Physical review letters}\ }\textbf {\bibinfo {volume} {121}},\
  \bibinfo {pages} {086803} (\bibinfo {year} {2018})}\BibitemShut {NoStop}%
\bibitem [{\citenamefont {Kunst}\ \emph {et~al.}(2018)\citenamefont {Kunst},
  \citenamefont {Edvardsson}, \citenamefont {Budich},\ and\ \citenamefont
  {Bergholtz}}]{kunst2018biorthogonal}%
  \BibitemOpen
  \bibfield  {author} {\bibinfo {author} {\bibfnamefont {F.~K.}\ \bibnamefont
  {Kunst}}, \bibinfo {author} {\bibfnamefont {E.}~\bibnamefont {Edvardsson}},
  \bibinfo {author} {\bibfnamefont {J.~C.}\ \bibnamefont {Budich}},\ and\
  \bibinfo {author} {\bibfnamefont {E.~J.}\ \bibnamefont {Bergholtz}},\
  }\bibfield  {title} {\bibinfo {title} {Biorthogonal bulk-boundary
  correspondence in non-hermitian systems},\ }\href@noop {} {\bibfield
  {journal} {\bibinfo  {journal} {Physical review letters}\ }\textbf {\bibinfo
  {volume} {121}},\ \bibinfo {pages} {026808} (\bibinfo {year}
  {2018})}\BibitemShut {NoStop}%
\bibitem [{\citenamefont {Edvardsson}\ \emph {et~al.}(2019)\citenamefont
  {Edvardsson}, \citenamefont {Kunst},\ and\ \citenamefont
  {Bergholtz}}]{edvardsson2019non}%
  \BibitemOpen
  \bibfield  {author} {\bibinfo {author} {\bibfnamefont {E.}~\bibnamefont
  {Edvardsson}}, \bibinfo {author} {\bibfnamefont {F.~K.}\ \bibnamefont
  {Kunst}},\ and\ \bibinfo {author} {\bibfnamefont {E.~J.}\ \bibnamefont
  {Bergholtz}},\ }\bibfield  {title} {\bibinfo {title} {Non-hermitian
  extensions of higher-order topological phases and their biorthogonal
  bulk-boundary correspondence},\ }\href@noop {} {\bibfield  {journal}
  {\bibinfo  {journal} {Physical Review B}\ }\textbf {\bibinfo {volume} {99}},\
  \bibinfo {pages} {081302} (\bibinfo {year} {2019})}\BibitemShut {NoStop}%
\bibitem [{\citenamefont {Bergholtz}\ \emph {et~al.}(2021)\citenamefont
  {Bergholtz}, \citenamefont {Budich},\ and\ \citenamefont
  {Kunst}}]{bergholtz2021exceptional}%
  \BibitemOpen
  \bibfield  {author} {\bibinfo {author} {\bibfnamefont {E.~J.}\ \bibnamefont
  {Bergholtz}}, \bibinfo {author} {\bibfnamefont {J.~C.}\ \bibnamefont
  {Budich}},\ and\ \bibinfo {author} {\bibfnamefont {F.~K.}\ \bibnamefont
  {Kunst}},\ }\bibfield  {title} {\bibinfo {title} {Exceptional topology of
  non-hermitian systems},\ }\href@noop {} {\bibfield  {journal} {\bibinfo
  {journal} {Reviews of Modern Physics}\ }\textbf {\bibinfo {volume} {93}},\
  \bibinfo {pages} {015005} (\bibinfo {year} {2021})}\BibitemShut {NoStop}%
\bibitem [{\citenamefont {Kunst}\ and\ \citenamefont
  {Dwivedi}(2019)}]{kunst2019non}%
  \BibitemOpen
  \bibfield  {author} {\bibinfo {author} {\bibfnamefont {F.~K.}\ \bibnamefont
  {Kunst}}\ and\ \bibinfo {author} {\bibfnamefont {V.}~\bibnamefont
  {Dwivedi}},\ }\bibfield  {title} {\bibinfo {title} {Non-hermitian systems and
  topology: A transfer-matrix perspective},\ }\href@noop {} {\bibfield
  {journal} {\bibinfo  {journal} {Physical Review B}\ }\textbf {\bibinfo
  {volume} {99}},\ \bibinfo {pages} {245116} (\bibinfo {year}
  {2019})}\BibitemShut {NoStop}%
\bibitem [{\citenamefont {Koch}\ and\ \citenamefont
  {Budich}(2020)}]{koch2020bulk}%
  \BibitemOpen
  \bibfield  {author} {\bibinfo {author} {\bibfnamefont {R.}~\bibnamefont
  {Koch}}\ and\ \bibinfo {author} {\bibfnamefont {J.~C.}\ \bibnamefont
  {Budich}},\ }\bibfield  {title} {\bibinfo {title} {Bulk-boundary
  correspondence in non-hermitian systems: stability analysis for generalized
  boundary conditions},\ }\href@noop {} {\bibfield  {journal} {\bibinfo
  {journal} {The European Physical Journal D}\ }\textbf {\bibinfo {volume}
  {74}},\ \bibinfo {pages} {1} (\bibinfo {year} {2020})}\BibitemShut {NoStop}%
\bibitem [{\citenamefont {Shen}\ \emph {et~al.}(2018)\citenamefont {Shen},
  \citenamefont {Zhen},\ and\ \citenamefont {Fu}}]{shen2018topological}%
  \BibitemOpen
  \bibfield  {author} {\bibinfo {author} {\bibfnamefont {H.}~\bibnamefont
  {Shen}}, \bibinfo {author} {\bibfnamefont {B.}~\bibnamefont {Zhen}},\ and\
  \bibinfo {author} {\bibfnamefont {L.}~\bibnamefont {Fu}},\ }\bibfield
  {title} {\bibinfo {title} {Topological band theory for non-hermitian
  hamiltonians},\ }\href@noop {} {\bibfield  {journal} {\bibinfo  {journal}
  {Physical review letters}\ }\textbf {\bibinfo {volume} {120}},\ \bibinfo
  {pages} {146402} (\bibinfo {year} {2018})}\BibitemShut {NoStop}%
\bibitem [{\citenamefont {Xue}\ \emph {et~al.}(2020)\citenamefont {Xue},
  \citenamefont {Wang}, \citenamefont {Zhang},\ and\ \citenamefont
  {Chong}}]{xue2020non}%
  \BibitemOpen
  \bibfield  {author} {\bibinfo {author} {\bibfnamefont {H.}~\bibnamefont
  {Xue}}, \bibinfo {author} {\bibfnamefont {Q.}~\bibnamefont {Wang}}, \bibinfo
  {author} {\bibfnamefont {B.}~\bibnamefont {Zhang}},\ and\ \bibinfo {author}
  {\bibfnamefont {Y.}~\bibnamefont {Chong}},\ }\bibfield  {title} {\bibinfo
  {title} {Non-hermitian dirac cones},\ }\href@noop {} {\bibfield  {journal}
  {\bibinfo  {journal} {Physical Review Letters}\ }\textbf {\bibinfo {volume}
  {124}},\ \bibinfo {pages} {236403} (\bibinfo {year} {2020})}\BibitemShut
  {NoStop}%
\bibitem [{\citenamefont {Zhu}\ \emph {et~al.}(2018)\citenamefont {Zhu},
  \citenamefont {Hou}, \citenamefont {Long}, \citenamefont {Chen},\ and\
  \citenamefont {Ren}}]{zhu2018simulating}%
  \BibitemOpen
  \bibfield  {author} {\bibinfo {author} {\bibfnamefont {W.}~\bibnamefont
  {Zhu}}, \bibinfo {author} {\bibfnamefont {S.}~\bibnamefont {Hou}}, \bibinfo
  {author} {\bibfnamefont {Y.}~\bibnamefont {Long}}, \bibinfo {author}
  {\bibfnamefont {H.}~\bibnamefont {Chen}},\ and\ \bibinfo {author}
  {\bibfnamefont {J.}~\bibnamefont {Ren}},\ }\bibfield  {title} {\bibinfo
  {title} {Simulating quantum spin hall effect in the topological lieb lattice
  of a linear circuit network},\ }\href@noop {} {\bibfield  {journal} {\bibinfo
   {journal} {Physical Review B}\ }\textbf {\bibinfo {volume} {97}},\ \bibinfo
  {pages} {075310} (\bibinfo {year} {2018})}\BibitemShut {NoStop}%
\bibitem [{\citenamefont {Serra-Garcia}\ \emph {et~al.}(2019)\citenamefont
  {Serra-Garcia}, \citenamefont {S{\"u}sstrunk},\ and\ \citenamefont
  {Huber}}]{serra2019observation}%
  \BibitemOpen
  \bibfield  {author} {\bibinfo {author} {\bibfnamefont {M.}~\bibnamefont
  {Serra-Garcia}}, \bibinfo {author} {\bibfnamefont {R.}~\bibnamefont
  {S{\"u}sstrunk}},\ and\ \bibinfo {author} {\bibfnamefont {S.~D.}\
  \bibnamefont {Huber}},\ }\bibfield  {title} {\bibinfo {title} {Observation of
  quadrupole transitions and edge mode topology in an lc circuit network},\
  }\href@noop {} {\bibfield  {journal} {\bibinfo  {journal} {Physical Review
  B}\ }\textbf {\bibinfo {volume} {99}},\ \bibinfo {pages} {020304} (\bibinfo
  {year} {2019})}\BibitemShut {NoStop}%
\bibitem [{\citenamefont {Ezawa}(2019{\natexlab{a}})}]{ezawa2019electric}%
  \BibitemOpen
  \bibfield  {author} {\bibinfo {author} {\bibfnamefont {M.}~\bibnamefont
  {Ezawa}},\ }\bibfield  {title} {\bibinfo {title} {Electric circuits for
  non-hermitian chern insulators},\ }\href@noop {} {\bibfield  {journal}
  {\bibinfo  {journal} {Physical Review B}\ }\textbf {\bibinfo {volume}
  {100}},\ \bibinfo {pages} {081401} (\bibinfo {year}
  {2019}{\natexlab{a}})}\BibitemShut {NoStop}%
\bibitem [{\citenamefont {Ezawa}(2019{\natexlab{b}})}]{ezawa2019non}%
  \BibitemOpen
  \bibfield  {author} {\bibinfo {author} {\bibfnamefont {M.}~\bibnamefont
  {Ezawa}},\ }\bibfield  {title} {\bibinfo {title} {Non-hermitian boundary and
  interface states in nonreciprocal higher-order topological metals and
  electrical circuits},\ }\href@noop {} {\bibfield  {journal} {\bibinfo
  {journal} {Physical Review B}\ }\textbf {\bibinfo {volume} {99}},\ \bibinfo
  {pages} {121411} (\bibinfo {year} {2019}{\natexlab{b}})}\BibitemShut
  {NoStop}%
\bibitem [{\citenamefont {Rafi-Ul-Islam}\ \emph {et~al.}(2021)\citenamefont
  {Rafi-Ul-Islam}, \citenamefont {Siu},\ and\ \citenamefont
  {Jalil}}]{rafi2021non}%
  \BibitemOpen
  \bibfield  {author} {\bibinfo {author} {\bibfnamefont {S.}~\bibnamefont
  {Rafi-Ul-Islam}}, \bibinfo {author} {\bibfnamefont {Z.~B.}\ \bibnamefont
  {Siu}},\ and\ \bibinfo {author} {\bibfnamefont {M.~B.}\ \bibnamefont
  {Jalil}},\ }\bibfield  {title} {\bibinfo {title} {Non-hermitian topological
  phases and exceptional lines in topolectrical circuits},\ }\href@noop {}
  {\bibfield  {journal} {\bibinfo  {journal} {New Journal of Physics}\ }\textbf
  {\bibinfo {volume} {23}},\ \bibinfo {pages} {033014} (\bibinfo {year}
  {2021})}\BibitemShut {NoStop}%
\bibitem [{\citenamefont {Haenel}\ \emph {et~al.}(2019)\citenamefont {Haenel},
  \citenamefont {Branch},\ and\ \citenamefont {Franz}}]{haenel2019chern}%
  \BibitemOpen
  \bibfield  {author} {\bibinfo {author} {\bibfnamefont {R.}~\bibnamefont
  {Haenel}}, \bibinfo {author} {\bibfnamefont {T.}~\bibnamefont {Branch}},\
  and\ \bibinfo {author} {\bibfnamefont {M.}~\bibnamefont {Franz}},\ }\bibfield
   {title} {\bibinfo {title} {Chern insulators for electromagnetic waves in
  electrical circuit networks},\ }\href@noop {} {\bibfield  {journal} {\bibinfo
   {journal} {Physical Review B}\ }\textbf {\bibinfo {volume} {99}},\ \bibinfo
  {pages} {235110} (\bibinfo {year} {2019})}\BibitemShut {NoStop}%
\bibitem [{\citenamefont {Imhof}\ \emph {et~al.}(2018)\citenamefont {Imhof},
  \citenamefont {Berger}, \citenamefont {Bayer}, \citenamefont {Brehm},
  \citenamefont {Molenkamp}, \citenamefont {Kiessling}, \citenamefont
  {Schindler}, \citenamefont {Lee}, \citenamefont {Greiter}, \citenamefont
  {Neupert} \emph {et~al.}}]{imhof2018topolectrical}%
  \BibitemOpen
  \bibfield  {author} {\bibinfo {author} {\bibfnamefont {S.}~\bibnamefont
  {Imhof}}, \bibinfo {author} {\bibfnamefont {C.}~\bibnamefont {Berger}},
  \bibinfo {author} {\bibfnamefont {F.}~\bibnamefont {Bayer}}, \bibinfo
  {author} {\bibfnamefont {J.}~\bibnamefont {Brehm}}, \bibinfo {author}
  {\bibfnamefont {L.~W.}\ \bibnamefont {Molenkamp}}, \bibinfo {author}
  {\bibfnamefont {T.}~\bibnamefont {Kiessling}}, \bibinfo {author}
  {\bibfnamefont {F.}~\bibnamefont {Schindler}}, \bibinfo {author}
  {\bibfnamefont {C.~H.}\ \bibnamefont {Lee}}, \bibinfo {author} {\bibfnamefont
  {M.}~\bibnamefont {Greiter}}, \bibinfo {author} {\bibfnamefont
  {T.}~\bibnamefont {Neupert}}, \emph {et~al.},\ }\bibfield  {title} {\bibinfo
  {title} {Topolectrical-circuit realization of topological corner modes},\
  }\href@noop {} {\bibfield  {journal} {\bibinfo  {journal} {Nature Physics}\
  }\textbf {\bibinfo {volume} {14}},\ \bibinfo {pages} {925} (\bibinfo {year}
  {2018})}\BibitemShut {NoStop}%
\bibitem [{\citenamefont {Lee}\ \emph {et~al.}(2018)\citenamefont {Lee},
  \citenamefont {Imhof}, \citenamefont {Berger}, \citenamefont {Bayer},
  \citenamefont {Brehm}, \citenamefont {Molenkamp}, \citenamefont {Kiessling},\
  and\ \citenamefont {Thomale}}]{lee2018topolectrical}%
  \BibitemOpen
  \bibfield  {author} {\bibinfo {author} {\bibfnamefont {C.~H.}\ \bibnamefont
  {Lee}}, \bibinfo {author} {\bibfnamefont {S.}~\bibnamefont {Imhof}}, \bibinfo
  {author} {\bibfnamefont {C.}~\bibnamefont {Berger}}, \bibinfo {author}
  {\bibfnamefont {F.}~\bibnamefont {Bayer}}, \bibinfo {author} {\bibfnamefont
  {J.}~\bibnamefont {Brehm}}, \bibinfo {author} {\bibfnamefont {L.~W.}\
  \bibnamefont {Molenkamp}}, \bibinfo {author} {\bibfnamefont {T.}~\bibnamefont
  {Kiessling}},\ and\ \bibinfo {author} {\bibfnamefont {R.}~\bibnamefont
  {Thomale}},\ }\bibfield  {title} {\bibinfo {title} {Topolectrical circuits},\
  }\href@noop {} {\bibfield  {journal} {\bibinfo  {journal} {Communications
  Physics}\ }\textbf {\bibinfo {volume} {1}},\ \bibinfo {pages} {1} (\bibinfo
  {year} {2018})}\BibitemShut {NoStop}%
\bibitem [{\citenamefont {Liu}\ \emph {et~al.}(2020)\citenamefont {Liu},
  \citenamefont {Ma}, \citenamefont {Yang}, \citenamefont {Zhang},
  \citenamefont {Gao}, \citenamefont {Xiang}, \citenamefont {Cui},\ and\
  \citenamefont {Zhang}}]{liu2020gain}%
  \BibitemOpen
  \bibfield  {author} {\bibinfo {author} {\bibfnamefont {S.}~\bibnamefont
  {Liu}}, \bibinfo {author} {\bibfnamefont {S.}~\bibnamefont {Ma}}, \bibinfo
  {author} {\bibfnamefont {C.}~\bibnamefont {Yang}}, \bibinfo {author}
  {\bibfnamefont {L.}~\bibnamefont {Zhang}}, \bibinfo {author} {\bibfnamefont
  {W.}~\bibnamefont {Gao}}, \bibinfo {author} {\bibfnamefont {Y.~J.}\
  \bibnamefont {Xiang}}, \bibinfo {author} {\bibfnamefont {T.~J.}\ \bibnamefont
  {Cui}},\ and\ \bibinfo {author} {\bibfnamefont {S.}~\bibnamefont {Zhang}},\
  }\bibfield  {title} {\bibinfo {title} {Gain-and loss-induced topological
  insulating phase in a non-hermitian electrical circuit},\ }\href@noop {}
  {\bibfield  {journal} {\bibinfo  {journal} {Physical Review Applied}\
  }\textbf {\bibinfo {volume} {13}},\ \bibinfo {pages} {014047} (\bibinfo
  {year} {2020})}\BibitemShut {NoStop}%
\bibitem [{\citenamefont {Goren}\ \emph {et~al.}(2018)\citenamefont {Goren},
  \citenamefont {Plekhanov}, \citenamefont {Appas},\ and\ \citenamefont
  {Le~Hur}}]{goren2018topological}%
  \BibitemOpen
  \bibfield  {author} {\bibinfo {author} {\bibfnamefont {T.}~\bibnamefont
  {Goren}}, \bibinfo {author} {\bibfnamefont {K.}~\bibnamefont {Plekhanov}},
  \bibinfo {author} {\bibfnamefont {F.}~\bibnamefont {Appas}},\ and\ \bibinfo
  {author} {\bibfnamefont {K.}~\bibnamefont {Le~Hur}},\ }\bibfield  {title}
  {\bibinfo {title} {Topological zak phase in strongly coupled lc circuits},\
  }\href@noop {} {\bibfield  {journal} {\bibinfo  {journal} {Physical Review
  B}\ }\textbf {\bibinfo {volume} {97}},\ \bibinfo {pages} {041106} (\bibinfo
  {year} {2018})}\BibitemShut {NoStop}%
\bibitem [{\citenamefont {Rafi-Ul-Islam}\ \emph {et~al.}(2020)\citenamefont
  {Rafi-Ul-Islam}, \citenamefont {Siu},\ and\ \citenamefont
  {Jalil}}]{rafi2020topoelectrical}%
  \BibitemOpen
  \bibfield  {author} {\bibinfo {author} {\bibfnamefont {S.}~\bibnamefont
  {Rafi-Ul-Islam}}, \bibinfo {author} {\bibfnamefont {Z.~B.}\ \bibnamefont
  {Siu}},\ and\ \bibinfo {author} {\bibfnamefont {M.~B.}\ \bibnamefont
  {Jalil}},\ }\bibfield  {title} {\bibinfo {title} {Topoelectrical circuit
  realization of a weyl semimetal heterojunction},\ }\href@noop {} {\bibfield
  {journal} {\bibinfo  {journal} {Communications Physics}\ }\textbf {\bibinfo
  {volume} {3}},\ \bibinfo {pages} {1} (\bibinfo {year} {2020})}\BibitemShut
  {NoStop}%
\bibitem [{\citenamefont {Luo}\ \emph {et~al.}(2018)\citenamefont {Luo},
  \citenamefont {Feng}, \citenamefont {Zhao},\ and\ \citenamefont
  {Yu}}]{luo2018nodal}%
  \BibitemOpen
  \bibfield  {author} {\bibinfo {author} {\bibfnamefont {K.}~\bibnamefont
  {Luo}}, \bibinfo {author} {\bibfnamefont {J.}~\bibnamefont {Feng}}, \bibinfo
  {author} {\bibfnamefont {Y.}~\bibnamefont {Zhao}},\ and\ \bibinfo {author}
  {\bibfnamefont {R.}~\bibnamefont {Yu}},\ }\bibfield  {title} {\bibinfo
  {title} {Nodal manifolds bounded by exceptional points on non-hermitian
  honeycomb lattices and electrical-circuit realizations},\ }\href@noop {}
  {\bibfield  {journal} {\bibinfo  {journal} {arXiv preprint arXiv:1810.09231}\
  } (\bibinfo {year} {2018})}\BibitemShut {NoStop}%
\bibitem [{\citenamefont {Hadad}\ \emph {et~al.}(2018)\citenamefont {Hadad},
  \citenamefont {Soric}, \citenamefont {Khanikaev},\ and\ \citenamefont
  {Alu}}]{hadad2018self}%
  \BibitemOpen
  \bibfield  {author} {\bibinfo {author} {\bibfnamefont {Y.}~\bibnamefont
  {Hadad}}, \bibinfo {author} {\bibfnamefont {J.~C.}\ \bibnamefont {Soric}},
  \bibinfo {author} {\bibfnamefont {A.~B.}\ \bibnamefont {Khanikaev}},\ and\
  \bibinfo {author} {\bibfnamefont {A.}~\bibnamefont {Alu}},\ }\bibfield
  {title} {\bibinfo {title} {Self-induced topological protection in nonlinear
  circuit arrays},\ }\href@noop {} {\bibfield  {journal} {\bibinfo  {journal}
  {Nature Electronics}\ }\textbf {\bibinfo {volume} {1}},\ \bibinfo {pages}
  {178} (\bibinfo {year} {2018})}\BibitemShut {NoStop}%
\bibitem [{\citenamefont {Ningyuan}\ \emph {et~al.}(2015)\citenamefont
  {Ningyuan}, \citenamefont {Owens}, \citenamefont {Sommer}, \citenamefont
  {Schuster},\ and\ \citenamefont {Simon}}]{PhysRevX.5.021031}%
  \BibitemOpen
  \bibfield  {author} {\bibinfo {author} {\bibfnamefont {J.}~\bibnamefont
  {Ningyuan}}, \bibinfo {author} {\bibfnamefont {C.}~\bibnamefont {Owens}},
  \bibinfo {author} {\bibfnamefont {A.}~\bibnamefont {Sommer}}, \bibinfo
  {author} {\bibfnamefont {D.}~\bibnamefont {Schuster}},\ and\ \bibinfo
  {author} {\bibfnamefont {J.}~\bibnamefont {Simon}},\ }\bibfield  {title}
  {\bibinfo {title} {Time- and site-resolved dynamics in a topological
  circuit},\ }\href {https://doi.org/10.1103/PhysRevX.5.021031} {\bibfield
  {journal} {\bibinfo  {journal} {Phys. Rev. X}\ }\textbf {\bibinfo {volume}
  {5}},\ \bibinfo {pages} {021031} (\bibinfo {year} {2015})}\BibitemShut
  {NoStop}%
\bibitem [{\citenamefont {Zhang}\ and\ \citenamefont
  {Franz}(2020)}]{PhysRevLett.124.046401}%
  \BibitemOpen
  \bibfield  {author} {\bibinfo {author} {\bibfnamefont {X.-X.}\ \bibnamefont
  {Zhang}}\ and\ \bibinfo {author} {\bibfnamefont {M.}~\bibnamefont {Franz}},\
  }\bibfield  {title} {\bibinfo {title} {Non-hermitian exceptional landau
  quantization in electric circuits},\ }\href
  {https://doi.org/10.1103/PhysRevLett.124.046401} {\bibfield  {journal}
  {\bibinfo  {journal} {Phys. Rev. Lett.}\ }\textbf {\bibinfo {volume} {124}},\
  \bibinfo {pages} {046401} (\bibinfo {year} {2020})}\BibitemShut {NoStop}%
\bibitem [{\citenamefont {Ezawa}(2019{\natexlab{c}})}]{PhysRevB.100.165419}%
  \BibitemOpen
  \bibfield  {author} {\bibinfo {author} {\bibfnamefont {M.}~\bibnamefont
  {Ezawa}},\ }\bibfield  {title} {\bibinfo {title} {Electric-circuit simulation
  of the schr\"odinger equation and non-hermitian quantum walks},\ }\href
  {https://doi.org/10.1103/PhysRevB.100.165419} {\bibfield  {journal} {\bibinfo
   {journal} {Phys. Rev. B}\ }\textbf {\bibinfo {volume} {100}},\ \bibinfo
  {pages} {165419} (\bibinfo {year} {2019}{\natexlab{c}})}\BibitemShut
  {NoStop}%
\end{thebibliography}%

\end{document}